\documentclass[a4paper,10pt]{article}
\usepackage[utf8]{inputenc}
\usepackage{amssymb, amsthm}
\usepackage{amsmath}
\usepackage{graphicx}
\graphicspath{ {Images/} }
\usepackage{cite}
\newtheorem{theorem}{Theorem}
\newtheorem{lemma}{Lemma}
\newtheorem{definition}{Definition}
\newtheorem{remark}{Remark}
\usepackage[toc,page]{appendix}
\begin{document}
\title{Improved Lower Bound on DHP: 
Towards the Equivalence of DHP and DLP for Important Elliptic Curves Used for Implementation}
\author{Prabhat Kushwaha \\
IISER Pune, Dr. Homi Bhabha Road, Pashan \\
Pune-411008, India \\
prabhat1412@gmail.com}
\bibliographystyle{plain}

\maketitle

\begin{abstract}
 In 2004, Muzereau \textit{et al.} showed how to use a reduction algorithm of the discrete logarithm problem to Diffie-
 Hellman problem in order to estimate lower bound on Diffie-Hellman problem on elliptic curves. They presented their 
 estimates for various elliptic curves that are used in practical applications. In this paper, we show that a much tighter
 lower bound for Diffie-Hellman problem on those curves can be achieved, if one uses the multiplicative group of a finite 
 field as an auxiliary group. Moreover, improved lower bound estimates on Diffie-Hellman problem for various recommended curves
 are also given which are the \textbf{tightest}; thus, leading us towards the equivalence of Diffie-Hellman problem and the 
 discrete logarithm problem for these recommended elliptic curves.
 \end{abstract}
\vskip 0.2 cm 
\textbf{Keywords}: Discrete Logarithm problem, lower bound of the Diffie-Hellman problem, elliptic curves used in practical
applications.
\section{Introduction}
 It is well known that the discrete logarithm problem(DLP) is one of two primitives that are commonly used as a building block in
 public key protocols, other being integer factorization. Computational difficulty in solving DLP is a security necessity for
 the protocols based on it. However, interesting thing about these DLP-based protocols is that, security of many of
 such protocols does not exactly rely on the hardness of DLP. For example, the ElGamal public key cryptosystem is secure if and 
 only if
 the Diffie-Hellman problem(DHP) is hard to solve \cite[Proposition 2.10]{sil}. That means, it is enough for an attacker to solve 
 DHP to break the ElGamal cryptosystem. The Diffie-Hellman key exchange, pairing-based cryptosystems, digital signature schemes and
 many more protocols are some other examples where the security of the protocol depends on hardness on DHP. This is why hardness
 of DHP is of utmost importance in public key cryptography.
 \vskip 0.5 cm
 If DLP is easy, DHP is easy because a solution of DLP immediately yields a solution of DHP. Therefore, the only meaningful 
 scenario to study the hardness of DHP is when DLP is known to be hard. Barring some weak elliptic curves over finite fields, 
 there are no efficient algorithm to solve the discrete logarithm problem on the group of points of an elliptic curve over 
 finite field (ECDLP) and thus, those elliptic curves are widely used for practical purposes. Thus, it is of paramount importance
 to study about the hardness of the elliptic curve Diffie-Hellman problem(ECDHP) from the point of view of practical cryptography. 
 The \textbf{central theme} of this paper is to study the hardness of ECDHP since a number of public key protocols are designed
 on such curves and their security depends only on the hardness of ECDHP.
 \vskip 0.5 cm
 \subsection{Summary of Existing work}
 To study the hardness of DHP the traditional method, and the only method known so far, involves reduction arguments from DLP to 
 DHP. In this section, we will summarize those reduction arguments. In such reductions, one tries to solve DLP efficiently(in 
 polynomial time of the input bit), using the existence of the a solution of DHP as sub-routine. If there exists such an 
 algorithm, we say that \textbf{DLP reduces to DHP} in polynomial time, denoted by DLP $\leq_{\textit{P}}$ DHP. Informally, 
 if DLP $\leq_{\textit{P}}$ DHP, it implies that DHP is at least as hard as DLP, or equivalently, DLP is no harder than DHP.
 Clearly, existence of any such reduction algorithm in case of elliptic curve groups would imply that ECDHP is hard, since 
 ECDLP is hard to solve.
 \vskip 0.5 cm
 The first, and the only, reduction algorithm known so far which reduces DLP to DHP was proposed by Maurer in his seminal  paper
 \cite{maurer1994towards}. He introduced the technique of implicit representation of elements of a finite field and indicated the 
 use of an auxiliary group in constructing such a reduction algorithm. Soon after that, Maurer and Wolf
 showed that DLP $\leq_{\textit{P}}$ DHP for any group $\mathbb{G}$ of prime order $p$; if we are able to find an elliptic
 curve over $\mathbb{F}_p$ with smooth order \cite{maurer1996difficulty, maurer1996diffie, maurer1999relationship,maurer2000diffie}.
 Smooth order of the auxiliary elliptic curve was the main reason behind the polynomial time reduction of DLP to DHP in their 
 algorithm because it ensures that the total number of group operations as well as the total number of calls to the DH-oracle
 required in their algorithm remain polynomial in the input size. However, it is exceptionally hard, in general, to construct 
 an elliptic curve over $\mathbb{F}_p$ of smooth order for large $p$, resulting in the failure of above theorem. Therefore, some
 alternate method was needed to study the hardness of DHP.
 \vskip 0.5 cm
 In 2004, Muzereau \textit{et al} re-visited the Maurer and Wolf reduction algorithm of DLP to DHP for special case of elliptic
 curve groups over finite field \cite{muzereau2004equivalence}. They explicitly constructed auxiliary elliptic curves, which
 were required in the reduction algorithm, for a number of elliptic curves recommended for practical implementation in 
 \textit{SEC 2}\cite{secg} by Standard for Efficient Cryptography Group(SECG) at Certicom Corporation. We will refer to those
 recommended elliptic curves in \textit{SEC 2} as \textbf{SECG curves}\cite{secg} throughout this paper.
 However, the orders of the auxiliary elliptic curves constructed were not smooth enough, making the cost of the reduction 
 algorithm exponential. Therefore, their reduction algorithm with those auxiliary elliptic curves failed to prove the polynomial
 reduction of ECDLP to ECDHP for those recommended SECG curves. Nevertheless, all was not lost as it might seem, since they were
 the first to give precise estimate of the number of group operations needed in such a
 reduction algorithm and showed how to use such a reduction algorithm to estimate the minimum number of group operations
 required to solve ECDHP on those SECG curves\cite[Table 1, Table 2]{muzereau2004equivalence}.  
 \vskip 0.2 cm
 Bentahar later applied the idea similar to Muzereau \textit{et al.} but constructed different auxiliary elliptic curves
 over $\mathbb{F}_p$ to improve those estimates of Muzereau \textit{et al.} on the exact lower bounds on ECDHP for those important
 SECG curves\cite[Table 1, Table 2]{bentahar2005equivalence}. 
 Since those lower bounds on ECDHP are assumed to be beyond the reach of present computational power, it establishes the security
 of several protocols relying on the hardness of ECDHP for those recommended SECG curves. This shows the significance of this
 remarkable approach.
\vskip 0.5 cm
\subsection{Our Contribution}
The algorithms of Muzereau and Bentahar both use the same reduction algorithm suggested by Maurer and Wolf. They used
suitable elliptic curves over a finite field as auxiliary groups. Our contribution in this paper is that a new reduction algorithm
of DLP to DHP is presented which uses, \textit{for the first time}, the multiplicative group of a finite field as an auxiliary 
group. Our reduction algorithm is also different from those used by Muzereau\cite{muzereau2004equivalence} and Bentahar
\cite{bentahar2005equivalence} or from any other previous reduction algorithm. Owing to this difference between our algorithm and
previous algorithms and the change in the auxiliary group from an elliptic curve over a finite field to the multiplicative group
of a finite field; our reduction algorithm requires very small number of DH-oracle calls. Our reduction algorithm results in 
increasing the lower bound on DHP, because the lower bound on DHP is inversely proportional to the number of calls to the 
DH-oracle. When applied to SECG curves studied first by Muzereau \textit{et al.} and then by Bentahar, our reduction algorithm 
improves the previous lower bounds on ECDHP.
\vskip 0.4 cm
More precisely, assuming that the best algorithm to solve DLP on an elliptic curve of order $p$ takes at 
least $\sqrt{p}$ group operations, Muzereau \textit{et al.} gave the following estimate on the lower bound on ECDHP \cite[Theorem 4]
{muzereau2004equivalence}:
 
 \begin{theorem}
  Let $p$ be a prime. Assuming in the interval $[p+1-\sqrt{p}, p+1 + \sqrt{p}]$ there is an integer which is product of three
  primes of roughly equal size, then there exists a string S which implies that the best algorithm to solve the ECDHP for an
  elliptic curve of order $p$ takes time at least $$ \textit{O}\left(\frac{\sqrt{p}}{ (log_2 p)^2 }\right)$$ group operations.
  \end{theorem}
 Under the same assumptions as above, our reduction algorithm will prove the following theorem which improves the lower bound
 on ECDHP in Theorem 1:
  \begin{theorem}
 For a prime $p$, assume that there exists a divisor $d$ of $p-1$ of the size roughly equal to $\sqrt[3]{p}$. Then, the best
 algorithm to solve ECDHP for an elliptic curve of order $p$ takes at least
 $$ \textit{O}\left(\frac{\sqrt{p}}{ log_2 d }\right)$$ group operations. 
 \end{theorem} 
  It is important to note that both the theorems above assume that the best known algorithm to solve ECDLP on an elliptic curve of order
 $p$ requires at least $\sqrt{p}$ group operations. 
 \vskip 0.3 cm
 Our result is significant as it applies to almost all the recommended SECG curves because such a divisor $d$ exists for almost
 all of those curves where prime $p$ is either the order of those elliptic curve groups or the largest prime divisor(with a very 
 small co-factor of either $2$ or $4$). 
 \vskip 0.2 cm
 Moreover, for curves SECP521R1, SECT409R1, SECT571R1, SECT571K1, Bentahar was unable to construct the auxiliary elliptic curves.
 However, we had no problem applying our algorithm to these curves and the lower bound estimates on these curves are also
 given here.
 \vskip 0.5 cm

\section{Notations and Definitions}
 Let $<\mathbb{G},+>$ be a cyclic(additive) group generated by $P$ and order of $P$ is a prime $p$.
 
 \begin{definition}
  Given $Q \in \mathbb{G}$, the problem of computing the integer $x$ modulo $p$ such that $Q = xP$ is called the 
  \textbf{discrete logarithm problem(DLP)} with respect to $P$.
 \end{definition}
\begin{definition}
 Given $ Q=xP, R=yP \in \mathbb{G}$($x, y$ are unknown integers), the problem of computing $S = xy P$ is called the
 \textbf{Diffie-Hellman problem(DHP)} with respect to $P$.
 \end{definition}
From the above definitions, it is clear that if one can compute $x$ from $Q=xP$ and then he can compute $xR = xyP \in\mathbb{G}$.
Thus, the solution of DLP readily yields the solution of DHP. However, as discussed earlier, we are interested in the reverse 
implication: does a solution of DHP solve DLP as well? To answer this question, reduction of DLP to DHP has been suggested and 
also given in some particular cases by Maurer and Wolf. As mentioned earlier, one tries to solve DLP assuming that a solution 
of DHP is known, or equivalently, one has access to a DH-oracle. We define it formally as follows:

\begin{definition}
  A \textbf{DH-oracle} is a function that takes $xP, yP \in \mathbb{G}$ as inputs and returns $xy P \in \mathbb{G}$ as output.
 We write it as $\mathcal{DH}(xP , yP) = xy P$.
\end{definition}

It was great insight of Maurer and Wolf who gave the first reduction algorithm that solved DLP using the DH-oracle as a sub-routine.
The algorithm used the idea of \textit{implicit representation of elements of a finite field $\mathbb{F}_p$} and 
\textit{auxiliary groups}.

\subsection{Implicit Representation of Elements of $\mathbb{F}_p$}
 
Let $\mathbb{G}$ be a cyclic group with generator $P$ whose order is a prime number $p$. Let $ y \in \mathbb{F}_p$. Then, 
$yP \in \mathbb{G}$ is called the \textit{implicit representation of $y\in \mathbb{F}_p$(with respect to $\mathbb{G}$ and $P$)}.
We denote this by $ y \rightsquigarrow yP$. 
\vskip 0.3 cm
Let $yP, zP \in \mathbb{G}$ be implicit representations of $y, z \in \mathbb{F}_p$ respectively. Then following basic algebraic 
operations in $\mathbb{F}_p$ can also be realized in $\mathbb{G}$ as follows:

\begin{itemize}
\item \textbf{Equality testing}: $ y = z$ if and only if $ yP = zP$.
\item \textbf{Addition}:  $y + z \rightsquigarrow yP + zP$ (1 group operation in $\mathbb{G}$).
\item \textbf{Subtraction}: $ y-z \rightsquigarrow yP -zP$(\textit{O}(log $p$) group operations in $\mathbb{G}$).
\item \textbf{Multiplication}: $ y \cdot z \rightsquigarrow {yz}P = {\mathcal{DH}}(yP, zP)$ (1 call to DH-oracle).
\item \textbf{Inversion}: $ y^{-1} = y^{p-2} = \underbrace{y \cdots y} \rightsquigarrow  {y^{p-2}}P$ (\textit{O}(log$_{2} p$) 
DH-oracle calls by using binary expansion). 
\end{itemize}

\vskip 0.2 cm	
Observe that the DH-oracle is used only for multiplication and inversion in $\mathbb{F}_p$. Therefore, number of 
DH-oracle calls required in the reduction algorithm increases with the increase in number of multiplication and inversions
in $\mathbb{F}_p$ required in the reduction algorithm. We will see the importance of this in later sections.

\subsection{Auxiliary Groups}
As the name suggests, any group(other than the group $\mathbb{G}$) is called an auxiliary group if it can be 
used to achieve the targeted goal of an algorithm. In the present context of DLP to DHP reduction, the goal is to solve DLP 
using the DH-oracle calls and implicit representations.
Therefore, two essential properties of a possible auxiliary group $\mathbb{H}$ are:
\begin{itemize}
 \item Elements of $\mathbb{H}$ can be represented as $m$-tuples of elements of $\mathbb{F}_p$ for some $m \geq 1$.
 \item Group operation in this auxiliary group $\mathbb{H}$ can be defined from algebraic operations in $\mathbb{F}_p$.
\end{itemize}
 \vskip 0.2 cm
 
These two necessary properties of $\mathbb{H}$ were suggested by Maurer and Wolf \cite{maurer1999relationship}. If $\mathbb{H}$ 
has these properties, then any computation in $\mathbb{H}$( for example, equality testing, exponentiation in $\mathbb{H}$) can also be
performed on their implicitly represented elements of $\mathbb{G}$. For more details, refer to \cite{maurer1999relationship}.
\vskip 0.2 cm
Moreover, Maurer and Wolf\cite{maurer1999relationship} also mentioned two classes of possible auxiliary groups, satisfying
above requirements: elliptic curves $\bar{E}(\mathbb{F}_p)$ and subgroups of $\mathbb{F}_{p^n}^{\times}$
for some $n\geq 1$. They called these groups \textit{applicable auxiliary groups over $\mathbb{F}_p$}. 

\subsection{General Idea of Solving DLP using Auxiliary Groups and Implicit Representation Computation}
 Let $\mathbb{G}$ be a cyclic group generated by $P$ and order of $P$ is a prime $p$. To solve DLP for $Q$, one requires to 
 find the integer $x$ where $Q = xP$. Moreover, we also have access to a DH-oracle on $\mathbb{G}$
 and we are allowed to make calls to the DH-oracle to solve DLP on $\mathbb{G}$. To this end, using auxiliary groups over
 $\mathbb{F}_p$ and computation on implicitly represented elements of $\mathbb{F}_p$, Maurer and Wolf gave the following general 
 idea for a DLP to DHP reduction algorithm:
 \begin{enumerate}
    \item Choose a cyclic auxiliary group $\mathbb{H}$ over $\mathbb{F}_p$ generated by $\zeta_0$.
  \item \textbf{Embed} the unknown $x$ into an implicitly represented element $c$ of $\mathbb{H}$.
  \item Compute discrete logarithm of $c$ with respect to $\zeta_0$ in $\mathbb{H}$ explicitly, using computation(in 
  $\mathbb{G}$) of implicitly represented elements of $\mathbb{F}_p$. Observe that computing implicit representations of finite 
  field elements is exactly the place where the DH-oracle is used.
  \item \textbf{Extract} the unknown $x$ from the discrete logarithm of $c$ with respect to $\zeta_0$ found in the last step.
 \end{enumerate}
\vskip 0.5 cm
It is interesting to note that all DLP to DHP reduction algorithms known so far are based on Maurer and Wolf's idea of 
implicit representations. More intriguing is the fact that as auxiliary groups, only elliptic curves over $\mathbb{F}_p$ of smooth 
order have been used and studied extensively. 
\vskip 0.3 cm
If we take $\mathbb{H} = \bar{E}(\mathbb{F}_p)$ as the auxiliary group with smooth order $N$ where elliptic curve $\bar{E}
(\mathbb{F}_p)$ is given by $Y^2 = X^3 + AX + B$; $A, B\in \mathbb{F}_p$ and generated by $P_0= (x_0, y_0)\in \mathbb{H}$, the 
reduction algorithm of Muzereau \textit{et al.}(which follows the above general idea) \textbf{embeds} the unknown $x$ implicitly into 
$c=Q_0=(x,y) \in \mathbb{H}$ for some $y\in \mathbb{F}_p$. After that, the discrete logarithm $k$ of $Q_0$ with respect to $P_0$ 
is computed \textit{explicitly} using computations on implicitly represented elements. The last step is to \textbf{extract} $x$ from $kP_0(=Q_0)$
which is the abscissa of the point $Q_0$. Observe that $Q_0 = (x, y)$ was not \textit{explicitly} known before the computation of $k$.
However, once we have $k$, we can compute $Q_0$ \textit{explicitly} using $P_0$ and $k$ as $Q_0 = kP_0$. 
Muzereau \textit{et al.} first computed $k$ modulo each prime power of $N$ by repeatedly applying Pohlig-Hellman algorithm on 
implicitly represented elements along with exhaustive search to find a collision, then used the Chinese Remainder Theorem to 
find $k$. Bentahar also applied the same method. We call this several instances of Pohlig-Hellman algorithm and exhaustive search in 
their reduction algorithms collectively as \textbf{sub-algorithm A}. For more details on this reduction, see\cite{muzereau2004equivalence}.
\vskip 0.6 cm
In the following section, we present a reduction algorithm that uses $\mathbb{F}_p^{\times}$ as an auxiliary group, instead of an 
elliptic curve over $\mathbb{F}_p$.

\section{DLP to DHP Reduction Algorithm using  $\mathbb{F}_p^{\times}$ as an Auxiliary Group}
 
The reduction algorithm presented here is an adaptation of Cheon's work used to solve DLPwAI\cite[Theorem 1]{cheon}. 
Cheon analyzed the security concerns on DLP given some additional(auxiliary) input. Much to our surprise, we found that his 
algorithm fits perfectly well into the general idea of Maurer and Wolf to reduce DLP to DHP using implicit 
representation with $\mathbb{F}_p^{\times}$ as an auxiliary group. 

The immediate and important application of this connection is that it gives us the tightest estimate known so far on the lower 
bound on ECDHP for those SECG curves. We present our DLP to DHP reduction algorithm using implicit representations and 
$\mathbb{F}_p^{\times}$ as auxiliary group in the following lemma:  
 
\begin{lemma}
Let $\mathbb{G}$ be a additive cyclic group generated by $P \in \mathbb{G}$ and the order of $P$ is a prime number $p$.
Let $ Q = xP \in \mathbb{G}$. Then, $x$ can be computed using at most $2$log$_2 p\left
(\left[\sqrt{\frac{p-1}{d}}\right]+{\left[\sqrt{d}\right]} \right)$ group operations and by making at most $2[$log$_2 d]$ 
calls to the DH-oracle. Here $d$ is a positive divisor of $p-1$ and $[.]$ is the greatest integer function.
\begin{proof}
As already discussed, the proof is based on implicit representation of elements of $\mathbb{F}_p$ using $\mathbb{H}=
\mathbb{F}_p^{\times}$ as the auxiliary group. Recall the unknown $x$ will be implicitly represented by $Q = xP \in \mathbb{G}$.
Furthermore, $\mathbb{F}_p^{\times}$ is a cyclic group with $\phi(p-1)$ generators, where $\phi$ is the Euler totient function. Since
a random element in $\mathbb{F}_p^{\times}$is a generator with probability
                   $$ \frac{\phi(p-1)}{p-1}> \frac{1}{6 log (log (p-1))} $$
which is large enough(see \cite{cheon}), it's easy to choose a generator of $\mathbb{F}_p^{\times}$. 
\vskip 0.3 cm 

Let $\zeta _ 0$ be a generator of $\mathbb{H}= \mathbb{F}_p^{\times}$, then 
\begin{equation}
x = \zeta_0 ^{i_0}   \, \,  (mod \, \, p) 
\end{equation}
for some integer $i_0$ such that $ 1 \leq i_0 \leq p-1$.
\vskip 0.3 cm
We want to compute $i_0$ \textit{explicitly} and then $x$ can be computed using above equation. Let $ \zeta = \zeta_0 ^d \, 
(mod \, p)$. Since $d | (p-1)$, there exists unique cyclic subgroup, $\mathbb{K}$ of $\mathbb{H}= \mathbb{F}_p^{\times}$ of 
order $\frac{p-1}{d}$, generated by $\zeta$. Now as $(x^d)^{\left(\frac{p-1}{d}\right)}=1$, it implies that 
$x^d \in \mathbb{K}$. Therefore, there exists unique non-negative integer $j$ with $1 \leq j \leq \left(\frac{p-1}{d}\right)$
such that
\begin{equation}
  x^d= \zeta ^j \, \, \, \, \, \, \, \, \, \,(mod \, p) \, 
\end{equation}
Let $d_1=\left[\sqrt{\frac{p-1}{d}}\right]$.  Since $j$ is between $1$ and $\frac{p-1}{d}$, 
there exist unique non-negative integers $u_1, v_1$ with $0 \leq u_1,v_1 \leq d_1$ such that $j=u_1 d_1-v_1$.
Plugging this value of $j$ in Equation 2, we get 
$$ x^d = \zeta ^{u_1  d_1} \zeta ^{-v_1} \, \, \, \, \, \, \, \, (mod \, p) \, $$ 
which implies,    
\begin{equation}
 \zeta^{v_1} x^d = (\zeta ^{d_1})^{u_1} \, \, \, \, \, \, \, \, (mod \, p) \, 
\end{equation}

Recall that equality of two field elements can also be checked on their implicitly represented elements as follows: 
$ y = z  \, \, $(\textit{in} $\mathbb{F}_p^{\times}$) \textit{is equivalent to  $yP = zP$ in $\mathbb{G}$}. Therefore, above
\textbf{implicit equation}  in $\mathbb{F}_p^{\times}$ is equivalent to following \textbf{explicit equation} in $\mathbb{G}$:
\begin{equation}
 \zeta ^{v_1}	( x^d P) = (\zeta ^{d_1})^{u_1} P 
\end{equation}

Since $x^d$ is multiplication by $x$ with itself $d$ times and we know $P$ and $x P$, we can compute implicit representation $x^d P$ 
of $x^d$, by making at most $2$[log$_2d$] calls to the DH-oracle, using a method similar to the double-and-add algorithm
\cite[Section 6.3]{sil}. 
\vskip 0.3 cm
Looking at Equation 4, it is clear that the elements on the left-hand side can be computed using $x^d P$ for any value of $v_1$ 
with  $0 \leq v_1 \leq d_1$, by repeated addition of previous terms by $\zeta$-times. Similarly, the elements on the right-hand side can
be computed for any value of $u_1$  with $0 \leq u_1 \leq d_1$ using $P$ by repeated addition of previous terms by $\zeta^{d_1}$-times.
So, we compute $\zeta^{v_1}(x^d P)$ for each $v_1$ with  $0 \leq v_1 \leq d_1$ and store them. Then, we compare them
with each of right-hand side terms(similar to Baby-Step Giant-Step(BSGS) algorithm \cite{menezes1996handbook}) to find a match
and it yields the integer $j= u_1 d_1-v_1$. 
\vskip 0.2 cm

Note that the non-negative integer $j=u_1 d_1-v_1$ in Equation 2 is nothing but $i_0$ modulo $\frac{p-1}{d}$. Now
to compute $i_0$ modulo ($p-1$) from this integer $j$, we apply division algorithm on $i_0$ with divisor $\frac{p-1}{d}$
to get a relation between $i_0$ and $j$ and it gives us, $i_0=\left(\frac{p-1}{d}\right)t+j$ for some non-negative integer $t$.
Observe that $0 \leq t < d$, otherwise $i_0 \geq p-1$, a contradiction. Therefore, the integer $t$ can be written uniquely as 
$t= {u_2}\left[\sqrt{d}\right] - v_2$ for some $0 \leq u_2, v_2 \leq \left[\sqrt{d}\right]$, again by the division algorithm. 
Thus, we get the following \textbf{implicit equation} in $\mathbb{H}=\mathbb{F}_p^{\times}$,	
\begin{equation}
 x = {\zeta_0}^{i_0} = {\zeta_0}^{j + t\left(\frac{p-1}{d}\right)} = {\zeta_0}^{j}  {\zeta_0}^{\left(\frac{p-1}{d}\right)
 \left({u_2}\left[\sqrt{d}\right] - v_2 \right)}
\end{equation}
which is equivalent to
\begin{equation}
 \left({\zeta_0}^{\frac{p-1}{d}}\right)^{v_2} x = \left({\zeta_0}^{\left(\frac{p-1}{d}\right)\left[\sqrt{d}\right]}\right)^{u_2} {\zeta_0}^{j}
\end{equation}
The last \textbf{implicit equation} in $\mathbb{H}=\mathbb{F}_p^{\times}$ is equivalent to the following \textbf{explicit 
equation} in $\mathbb{G}$,
\begin{equation}
\left({\zeta_0}^{\frac{p-1}{d}}\right)^{v_2} (xP) = \left({\zeta_0}^{\left(\frac{p-1}{d}\right)\left[\sqrt{d}\right]}\right)^{u_2} ({\zeta_0}^{j} P)
\end{equation}

As $xP$ and ${\zeta_0}^{j} P_0$ are known, we can solve for $u_2, v_2$ by finding a match between two sides of the Equation 7 
using BSGS algorithm. This solution for $u_2, v_2$ would give us, 
$$i_0 = \left(\frac{p-1}{d}\right) \left({u_2}\left[\sqrt{d}\right]- v_2 \right) +j$$
Thus, we have \textbf{explicitly} computed $i_0$. 
Lastly, we \textbf{extract} the original discrete logarithm $x$ from this $i_0$ and the relation $x = {\zeta_0}^{i_0}$. 

\vskip 0.3 cm
It is easy to see that it takes at most $2$log$_2 p\left(\left[\sqrt{\frac{p-1}{d}} \right]\right)$ group operations to find a match 
in Equation 4  and at most $2$log$_2 p\left(\left[\sqrt{d}\right]\right)$ group operations to find a match in Equation 7. 
Therefore, we have computed the discrete logarithm $x$ using at most $2$log$_2 p\left(\left[\sqrt{\frac{p-1}{d}}\right] +
{\left[\sqrt{d}\right]}\right)$ group operations and by making at most $ 2$[log$_2 d$] calls to the DH-oracle. 
This completes the proof of the lemma.
\end{proof}
\end{lemma} 

\begin{remark}
 One can get rid of the factor log$_2 p$ from the above time complexity using KKM improvement \cite{kkm}. Then, above time
 complexity reduces to $2\left(\left[\sqrt{\frac{p-1}{d}} \right] +{\left[\sqrt{d}\right]}\right)$. 
\end{remark}

\begin{remark}
Observe that $x^d$ is unknown in Equation 3 because $x$ is unknown. This makes Equation 3 an \textbf{implicit equation} in 
$\mathbb{H}=\mathbb{F}_p^{\times}$. This is exactly the place where implicit representation computation comes into play, to
compute implicit representation $x^d P$ of $x^d$. 
Moreover, to compute $i_0$ modulo $(p-1)$ from the integer $j$, the idea used in our algorithm is from 
\cite[Theorem 1]{cheon} which uses the division algorithm on integers along with BSGS algorithm on implicitly represented 
elements. In all, we have used Pohlig-Hellman algorithm once, BSGS algorithm twice in our reduction algorithm. We call this 
single occurrence of Pohlig-Hellman algorithm and the use of BSGS  algorithm twice in our reduction algorithm, collectively as
\textbf{sub-algorithm B}.
\end{remark}
 
\begin{remark}
Our algorithm follows the general idea of DLP to DHP reduction algorithm by Maurer and  Wolf with $\mathbb{H}=\mathbb{F}_p^{\times}$
as the auxiliary group where the unknown $x$ is \textit{embedded implicitly} into itself, i.e. $c =x$. To the best of our
knowledge, our algorithm is the \textbf{first DLP to DHP reduction algorithm} that uses $\mathbb{H}=\mathbb{F}_p^{\times}$ as 
an auxiliary group but does not use the Chinese Remainder Theorem to compute the discrete logarithm.
\end {remark}

\section{Main Results}
As stated earlier, to prove computational equivalence of DLP and DHP on a group $\mathbb{G}$ of prime order $p$, one needs to construct an elliptic
curve over $\mathbb{F}_p$ of smooth order. This is an exceptionally hard task. Therefore, one has to look for some 
alternative ways to measure the hardness of DHP. The next best thing to the computation equivalence of DLP and DHP would be to
somehow estimate the minimum number of group operations required to solve DHP. That is exactly what Muzereau \textit{et al.}
\cite{muzereau2004equivalence} did for the elliptic curves groups recommended for practical implementation by SECG\cite{secg}.
Their idea was to  construct a DLP to DHP reduction algorithm in which total number of group operations needed in the 
reduction algorithm should be insignificant when compared to the cost of solving DLP. Once we have such a reduction algorithm, 
they proposed that 
ratio of the cost of DLP and the number of calls to the DH-oracle needed in the algorithm gives the minimum number of group operations
any algorithm that breaks DHP would require. 
\vskip 0.4 cm

\subsection{Proof of Theorem 2}
\begin{proof}
Since we are dealing with an elliptic curve of prime order $p$, we assume that the best algorithm to solve elliptic curve 
discrete logarithm problem will take at least $\sqrt{p}$  group operations. First, we give the general set up  of estimating the
lower bound on DHP. 
\vskip 0.3 cm
Let $C_{DLP}$, $C_{DHP}$ denote the time complexity of solving DLP and DHP respectively. Therefore, in the view of
a general DLP to DHP reduction algorithm, we get $C_{DLP} = n \cdot C_{DHP} + M$ where $n$ is the number of calls to the DH- oracle 
and $M$ is the number of group operations required in the reduction algorithm. Now, if we assume that $M \ll C_{DLP}$, then we 
have:
$$ C_{DHP} = \frac{C_{DLP} - M}{n} \approx \frac{C_{DLP}}{n} $$ 
If we set $ T_{DH} = \frac{C_{DLP}}{n}$, the number \textbf{$T_{DH}$} is exactly what gives the minimum number of group 
operations needed by any algorithm to solve DHP, assuming $ M \ll C_{DLP}$. This is how Muzereau \textit{et al.}
\cite{muzereau2004equivalence} estimated the
minimum number of group operations required by any algorithm that solves DHP. Of course, the aim would be to make $n$ as small as possible
to have the value of $T_{DH}$ as large as possible.
\vskip 0.4 cm

Now, we prove the lower bound on ECDHP. In case of $\mathbb{G}$ being an elliptic curve group of prime order $p$,
one can take $C_{ECDLP} = \sqrt{p}$ under our assumption. Since
there is a divisor $d$ of $p-1$ such that $d \approx \sqrt[3]{p}$, then it is easy to check that
$M \leq 2\left(\left[\sqrt{\frac{p-1}{d}} \right] +{\left[\sqrt{d}\right]} \right)\approx\sqrt[3]{p}$, satisfying the
condition $ M \ll C_{ECDLP}=\sqrt{p}$. Since $n \leq 2$[log$_2d$], we finally get, 
$$T_{DH}=\textit{O}\left(\frac{\sqrt{p}}{ log_2 d }\right).$$
This implies that the minimum number group operation to solve ECDHP on any elliptic curve group of prime order $p$ by any 
algorithm is of the order of $\textit{O}\left(\frac{\sqrt{p}}{ log_2 d}\right)$ if there exists a divisor $d$ of ($p-1$) of size
approximately $\sqrt[3]{p}$. This completes the proof.
\end{proof}
\begin{remark}
If we assume that a divisor $d$ of $p-1$ of size approximately $\sqrt[3]{p}$ exists, then the above result shows that the cost of
ECDHP is getting closer to the cost ECDLP.
\end{remark}
\begin{remark}
Note that the total number of group operations, $M$ needed in the reduction algorithms of Muzereau \textit{et al.}
\cite{muzereau2004equivalence}
and Bentahar \cite{bentahar2005equivalence} was also of the same order i.e. $ M \approx \sqrt[3]{p}$. This 
indicates the importance of such a divisor $d$ of size approximately $\sqrt[3]{p}$ in our reduction algorithm.
\end{remark}

\begin{remark}
 While the above reduction algorithm(with $\mathbb{F}_p^{\times}$ as an auxiliary group) as well as the previous reduction
 algorithms(with $\bar{E}(\mathbb{F}_p)$ as an auxiliary group) both depend on Maurer and Wolf implicit representation
 computation, it is clear that the \textbf{sub-algorithm B} used in our reduction algorithm and \textbf{sub-algorithm A} used in 
 previous reduction algorithms are quite different. Moreover, our reduction algorithm uses division algorithm to compute the 
 discrete logarithm while using \textbf{sub-algorithm B}. On the other hand, in previous reduction algorithms
 (which have $\bar{E}(\mathbb{F}_p)$ as an auxiliary group), the Chinese Remainder Theorem was used to compute the discrete 
 logarithm using \textbf{sub-algorithm A}.
 \end{remark}

\subsection{Improved value of $T_{DH}$ : Advantage of $\mathbb{F}_p^{\times}$ over $\bar{E}(\mathbb{F}_p) $}
The difference between \textit{sub-algorithm A} and \textit{sub-algorithm B} as well as the change of auxiliary group from 
$\bar{E}(\mathbb{F}_p)$ to $\mathbb{F}_p^{\times}$ both have their implications on the number of DH-oracle calls, consequently 
affecting the value of $T_{DH}$. Since \textit{Sub-algorithm A} used in previous reduction algorithms required several iterations
of Pohlig-Hellman algorithm, one had to compute a large number of implicitly represented elements in those reduction algorithms.
Therefore, a large number of DH-oracle calls were needed in the previous reduction algorithms. On the other hand, our reduction algorithm
while using  \textit{sub-algorithm B} requires \textbf{only} one implicitly represented element $x^d P $ of 
$x^d\in\mathbb{F}_p^{\times}$. This element can be computed by using at most $n\leq 2$[log$_2d$] DH-oracle calls
which can further be made really small by taking small value of $d$.
\vskip 0.2 cm
Recall that addition operation in $\bar{E}(\mathbb{F}_p)$ requires many multiplications in $\mathbb{F}_p$(one multiplication in 
$\mathbb{F}_p$ means one DH-oracle call to compute implicit representation) and many inversions in $\mathbb{F}_p$
(one inversion in $\mathbb{F}_p$ means on average $\frac{3}{2}$[log$_2 p$] calls to the DH-oracle to compute implicit 
representation). Thus, in terms of DH-oracle calls, computing the sum of elements in $\bar{E}(\mathbb{F}_p)$ is much more
expensive than multiplying elements in $\mathbb{F}_p^{\times}$.
\vskip 0.3 cm
Since our main aim through this reduction algorithm is to increase the value of $T_{DH}$ which is inversely proportional to 
number of DH-oracle calls $n$, it will be nice to reduce the number of DH-oracle calls as much as possible.
That is exactly what our reduction algorithm does using \textit{sub-algorithm B} and $\mathbb{F}_p^{\times}$ as the auxiliary group.
This shows that the advantage of our reduction algorithm over previous reduction algorithms which used \textit{sub-algorithm A}
and $\mathbb{H} = \bar{E}(\mathbb{F}_p)$ as auxiliary groups, for getting improved value of $T_{DH}$.

\subsection{Improved Lower Bound on ECDHP for SECG curves}

In this section, we study about the lower bound on ECDHP for various important elliptic curves parameters\cite{secg} and 
show the improvement made by our reduction algorithm on the lower bound on ECDHP for those curves. These curves are 
recommended in SEC 2 by Standard of Efficient Cryptography Group(SECG) at Certicom Corporation to be used for practical 
purposes and we have been calling those curves SECG curves. 
These SECG curves are divided into two sub-categories: curves over prime fields of large odd characteristic and curves over 
binary fields. The prime $p$ denotes the order of those SECG curves defined over prime fields of odd characteristic. For 
remaining SECG curves defined over binary fields, $p$ denotes the prime divisor of the order of the curve, with a very
small co-factor of either $2$ or $4$.     
\vskip0.2 cm
It should also be noted that SECG curves~\cite{secg} include all curves recommended by NIST~\cite{nist} and the most used 
ones in ANSI~\cite{ansi}. These covers the most commonly used curves in practice. Thus, these are important curves from the 
point of view of public key cryptography.  
\vskip 0.2 cm
Muzereau \textit{et al.}~\cite{muzereau2004equivalence} used the value of $T_{DH}$ as the lower bound on group operations 
to break DH-protocol and also gave the estimates for $T_{DH}$ on various SECG curves. Thereafter, Bentahar~\cite{bentahar2005equivalence} 
improved the previous values of $T_{DH}$ given by Muzereau\textit{et al.} and his estimates remain the best estimates till 
date.

\vskip 0.3 cm
Now, in our algorithm, with $\mathbb{F}_p^{\times}$ as the auxiliary group, we have $n\leq2$[log$_2d$] and
$M\leq 2\left(\left[\sqrt{\frac{p-1}{d}} \right] +{\left[\sqrt{d}\right]}\right)$ where $d$ is some divisor of $p-1$. 
As per the discussion above, to achieve a tighter(larger) value of $T_{DH}$ using our reduction algorithm, one should try to make
$n\leq2$[log$_2d$] as small as possible, which forces $d$ to be small as well. On the other hand, we have to make sure that $M\approx
\sqrt[3]{p}$, so that it does not violate $M \ll C_{ECDLP} = \sqrt{p}$. It is not hard to see that for really small value of $d$,
$M$ is inversely proportional to $d$. Therefore, too small value of $d$ must not be used to avoid the violation of $M \ll C_{ECDLP}
=\sqrt{p}$. Also note that $d \approx \sqrt[3]{p}$ yields $M \approx \sqrt[3]{p}$ in our reduction algorithm. 
\vskip 0.4 cm

Keeping all these in mind, we factored $p-1$ and found that most of SECG curves contain divisors $d$ which are between $\sqrt[3]{p}$
and $\sqrt{p}$ and we have taken the smallest such $d$ in the range $\sqrt[3]{p}$ and $\sqrt{p}$ to compute the values in Table 1 and
2 given below. For those curves where such a divisor $d$ does not exist, we have chosen the largest $d$ less than $\sqrt[3]{p}$
to compute the values in the tables.

\vskip 0.3 cm
For those choices of $d$, we calculated \textit{exact} number of the DH-oracle calls, $n \leq 2$[log$_2 d$] using binary expansion 
of $d$. The values of $n$ thus achieved are significantly small as compared with the values of $n$ shown by Bentahar
\cite{bentahar2005equivalence}(and much smaller than those in the work of Muzereau \textit{et al.}\cite{muzereau2004equivalence}). 
Consequently, these significantly small values of $n$ resulted in much tighter(larger)values of $T_{DH}$ for all SECG curves.
Therefore, it implies that we have given the \textbf{tightest lower bound}, known so far, on ECDHP for all SECG curves\cite{secg}
(except SECP224K1). In other words, our results shows the gap between the cost of ECDHP and ECDHP to be the \textbf{least}(known
so far) for these curves and it leads us one step closer towards the computational equivalence of ECDHP and ECDLP for these important
curves.
\vskip 0.3 cm
One additional advantage of our algorithm is that the values of $M$ in our algorithm are less than or of almost same order
as the ones given by Bentahar~\cite{bentahar2005equivalence} for most of SECG curves. 

\pagebreak

 \textbf{Table 1} : Summary of results for curves of large prime characteristic
$$\begin{tabular}{|c|c|c|c|c|c|c|}
\hline  &  SECP Curve & log$_2 \sqrt{|E|}$ & log$_2 M$ & log$_2 n$ &  log$_2  T_{\textit{DH}}$ & \textit{ADV}\\
\hline  &  SECP112R1 & 55.89 & 48.34 & 4.59 & 51.30 & 6.90 \\
\hline  &  SECP112R2 & 54.90 & 37.54 & 5.88 & 49.01 & 5.51 \\
\hline  &  SECP128R1 & 64.00 & 43.45 & 6.02 & 57.98 & 5.58 \\
\hline  &  SECP128R2 & 63.00 & 48.23 & 5.49 & 57.51 & 6.11 \\
\hline  &  SECP160K1 & 80.00 & 48.39 & 6.55 & 73.45 & 5.45 \\
\hline  &  SECP160R1 & 80.00 & 53.85 & 6.30 & 73.70 & 5.70 \\
\hline  &  SECP160R2 & 80.00 & 47.53 & 6.70 & 73.30 & 5.30 \\
\hline  &  SECP192K1 & 96.00 & 84.31 & 5.36 & 90.64 & 6.84 \\
\hline  &  SECP192R1 & 96.00 & 55.51 & 6.97 & 89.03 & 5.23 \\
\hline  &  SECP224R1 & 112.00 & 98.50 & 5.55 & 106.45 & 6.85 \\
\hline  &  SECP224K1 & - & - & - & - & - \\
\hline  &  SECP256K1 & 128.00 & 86.12 & 7.00 & 121.00 & 5.60 \\
\hline  &  SECP256R1 & 128.00 & 86.06 & 7.00 & 121.00 & 5.60 \\
\hline  &  SECP384R1 & 192.00 & 141.33 & 7.33 & 184.67 & 5.87 \\
\hline  &  SECP521R1 & 260.50 & 196.26 & 7.67 & 252.83 & 6.03\\
\hline
\end{tabular}$$

Table 1 and Table 2 present the key values, log$_2 M$, log$_2 n$ and log$_2 T_{DH}$ for various SECG curves. The 
tables also have the value of log$_2 \sqrt{|E|}$ which refers to the assumed minimum cost of solving DLP in that particular SECG 
curve $E$. The column under \textit{ADV} shows the number of security bits gained by the values of $T_{DH}$ in our algorithm over
the previous best known
values of $T_{DH}$ given by Bentahar\cite{bentahar2005equivalence}. Moreover, the present algorithm works for the curves SECP521R1,
SECT571R1, SECT571K1 as well which were out of reach in previous work due to inability to construct auxiliary
elliptic curves,  and Tables 1 and Table 2  give the key data for these curves as well.
\vskip 0.2 cm
It should also be remarked that the current algorithm fails for the curve SECP224K1 as there does not exist any divisor of
$p-1$ of appropriate size. Therefore, Bentahar's result still gives the tightest value of $T_{DH}$ for this curve.
\vskip 0.3 cm
To understand the advantage gained by our result over the work of Bentahar\cite{bentahar2005equivalence}, as an example we consider 
the security of ECDHP for SECP256R1. The best known algorithm at present to solve ECDLP on this curve takes on an average
$2^{128}$ group operations. Now, our algorithm implies that ECDHP can not be solved in less than $2^{121.00}$ group 
operations, in contrast to $2^{115.40}$ group operations from the work of Bentahar \cite{bentahar2005equivalence}. This shows that 
there is a gain factor of $2^{5.60}$ over the previous best known result given by Bentahar for 
the curve SECP256R1, see Table 1. If we assume that today's computational power is incapable of performing $2^{121.00}$ group
operations(which is considered to be true by many), then ECDHP on the curve SECP256R1 is secure and any cryptography 
protocol which rely on DHP for its security can safely be implemented on the curve SECP256R1, under the assumption above.

\, \, \,  \textbf{Table 2.} Summary of results for curves of even characteristic
$$\begin{tabular}{|c|c|c|c|c|c|c|}
\hline  &  SECT Curve & log$_2 \sqrt{|E|}$ & log$_2 M$ & log$_2 n$ & log$_2 T_{\textit{DH}}$ & \textit{ADV} \\ 
\hline   & SECT113R1 & 56.00 & 38.06 & 5.67 & 50.33 & 5.73 \\
\hline  &  SECT113R2 & 56.00 & 38.17 & 5.76 & 50.25 & 5.65 \\
\hline  &  SECT131R1 & 65.00 & 58.75 & 4.46 & 60.54 & 7.24 \\
\hline  &  SECT131R2 & 65.00 & 51.57 & 5.43 & 59.57 & 6.27 \\
\hline  &  SECT163K1 & 81.00 & 54.56 & 6.36 & 74.64 & 5.64 \\
\hline  &  SECT163R1 & 81.00 & 54.69 & 6.36 & 74.64 & 5.64 \\
\hline  &  SECT163R2 & 81.00 & 67.16 & 5.56 & 75.45 & 6.45 \\
\hline  &  SECT193R1 & 96.00 & 61.74 & 6.76 & 89.25 & 5.45 \\
\hline  &  SECT193R2 & 96.00 & 56.08 & 6.99 & 89.01 & 5.21 \\
\hline  &  SECT233K1 & 115.50 & 79.89 & 6.77 & 108.73& 5.73 \\
\hline  &  SECT233R1 & 116.00 & 77.72 & 6.92 & 109.08 & 5.58 \\
\hline  &  SECT239K1 & 116.00 & 79.70 & 6.87 & 111.63 & 5.63 \\
\hline  &  SECT283K1 & 140.50 & 94.51 & 7.15& 133.35 & 5.65 \\
\hline  &  SECT283R1 & 141.00 & 94.61 & 7.18 & 133.82 & 5.62 \\
\hline  &  SECT409K1 & 203.50 & 150.09 & 7.44 & 196.07 & 5.87 \\
\hline  &  SECT409R1 & 204.00 & 136.70 & 7.66 & 196.34 & 5.64 \\
\hline  &  SECT571K1 & 284.50 & 190.46 & 8.08 & 276.41 & 5.71 \\
\hline  &  SECT571R1 & 285.00 & 190.77 & 8.15 & 276.85 & 5.65 \\
\hline
\end{tabular}$$

\vskip 0.3 cm

\section{Conclusion} 
In this paper, we have presented first ever DLP to DHP reduction algorithm on a group $\mathbb{G}$ of prime order $p$ using 
$\mathbb{F}_p^{\times}$ as an auxiliary group in the implicit representation method. Earlier work used elliptic curves over 
$\mathbb{F}_p$ as auxiliary groups. We also established the advantage of our reduction algorithm over previously known reduction
algorithms to achieve better(increased) lower bound on the number of operations needed to solve DHP. As a consequence 
of our reduction algorithm, we have presented the \textbf{tightest lower bound} known
so far on ECDHP for all recommended SECG curves\cite{secg}(except SECP224K1). This work is of practical significance as it provides
tighter security for protocols which depend on ECDHP for their security.  Moreover, it leads us towards the computational 
equivalence of DHP and DLP for these SECG curves since the gap between the cost of DHP and DLP has been further reduced for
these curves.
\vskip 0.6 cm
\section*{Acknowledgement}
\vskip 0.2 cm
I wish to thank my advisor Dr. Ayan Mahalanobis for his continuous help and excellent guidance throughout this project.

\bibliography{Arxiv}
 
\end{document}